\documentclass[twocolumn]{aastex631}

\newcommand{\macc}{$\dot{M}_\mathrm{acc}$}

\shorttitle{Photometric Diagnostics of Magnetospheric Accretion Geometry in YSOs}
\shortauthors{Venuti et al.}

\begin{document}

\title{Preparing for the Early eVolution Explorer: Photometric Diagnostics of Magnetospheric Accretion Geometry in Young Stellar Objects}

\author[0000-0002-4115-0318]{Laura Venuti}
\affiliation{SETI Institute, 339 Bernardo Ave., Suite 200, Mountain View, CA 94043, USA}
\affiliation{Visiting Fellow, School of Physics, UNSW Science, Kensington, NSW 2052, Australia}
\correspondingauthor{Laura Venuti} \email{lvenuti@seti.org}

\author[0000-0003-1639-510X]{Connor E. Robinson}
\affiliation{Physics and Astronomy Division, Alfred University, 1 Saxon Drive, Alfred, NY 14802, USA}

\author[0000-0002-3656-6706]{Ann Marie Cody}
\affiliation{SETI Institute, 339 Bernardo Ave., Suite 200, Mountain View, CA 94043, USA}

\author[0000-0002-5258-6846]{Eric Gaidos}
\affiliation{Department of Earth Sciences, University of Hawai'i at M\={a}noa, Honolulu, HI 96822, USA}
\affiliation{Department of Astrophysics, University of Vienna, T\"urkenschanzstrasse 17, 1180 Vienna, Austria}

\author[0000-0001-7891-8143]{Meredith A. MacGregor}
\affiliation{Department of Physics \& Astronomy, John Hopkins University, 3400 N. Charles Street, Baltimore, MD 21218, USA}

\author[0000-0001-8292-1943]{Neal J. Turner}
\affiliation{SETI Institute, 339 Bernardo Ave., Suite 200, Mountain View, CA 94043, USA}

\author[0009-0001-4487-7299]{Mark R. Swain}
\affiliation{NASA Jet Propulsion Laboratory, California Institute of Technology, 4800 Oak Grove Drive, Pasadena, CA 91109, USA}

\author[0000-0002-4891-3517]{George Zhou}
\affiliation{Centre for Astrophysics, University of Southern Queensland, West Street, Toowoomba, QLD 4350, Australia}

\author[0000-0001-9158-9276]{Sydney Vach}
\affiliation{Centre for Astrophysics, University of Southern Queensland, West Street, Toowoomba, QLD 4350, Australia}
\affiliation{European Southern Observatory, Karl-Schwarzschild-Str. 2, 85748 Garching bei M\"unchen, Germany}

\author[0000-0003-2515-4623]{Lukas Gehrig}
\affiliation{Department of Astrophysics, University of Vienna, T\"urkenschanzstrasse 17, 1180 Vienna, Austria}

\author[0000-0002-8828-6386]{Christopher M. Johns-Krull}
\affiliation{Department of Physics and Astronomy, Rice University, 6100 Main Street, Houston, TX 77005, USA}

\author[0000-0003-3616-6822]{Zhaohuan Zhu}
\affiliation{Department of Physics and Astronomy, University of Nevada, Las Vegas, 4505 S. Maryland Pkwy, Las Vegas, NV 89154, USA}
\affiliation{Nevada Center for Astrophysics, University of Nevada, Las Vegas, 4505 S. Maryland Pkwy, Las Vegas, NV 89154, USA}

\author[0000-0002-0040-6815]{Jennifer A. Burt}
\affiliation{NASA Jet Propulsion Laboratory, California Institute of Technology, 4800 Oak Grove Drive, Pasadena, CA 91109, USA}

\author[0000-0002-0583-0949]{Ward S. Howard}
\affiliation{Department of Astrophysical and Planetary Sciences, University of Colorado, 2000 Colorado Avenue, Boulder, CO 80309, USA}

\author[0000-0002-2085-7402]{Damon Landau}
\affiliation{NASA Jet Propulsion Laboratory, California Institute of Technology, 4800 Oak Grove Drive, Pasadena, CA 91109, USA}

\author[0000-0002-7260-5821]{Evgenya Shkolnik}
\affiliation{School of Earth and Space Exploration, Arizona State University, 781 Terrace Mall, Tempe, AZ 85287, USA}

\begin{abstract}
The inner disk truncation radius, $R_T$, plays a crucial role in the regulation of star-disk interaction and the early evolution of star-disk-planet systems; however, measuring this parameter is observationally challenging. We present a new method for determining $R_T$ in young accreting systems that hinges on the color dependence of the accretion shock emission in multi-band time-domain surveys. Based on the accretion simulations of \citet{robinson2017,robinson2021}, we produce synthetic color-magnitude diagrams at near-UV and optical wavelengths that predict the loci of accreting stars as a function of $R_T$. We test these model predictions on young stars with interferometric $R_T$ estimates, finding very good agreement in our results. We apply this novel technique to a pilot survey of 26 classical T Tauri stars in Taurus and Upper Scorpius. We find a predominance of sources with small truncation radii, $R_T < 4\ R_\star$, and an overall distribution of $R_T$ that is statistically similar to that inferred from interferometric studies, while differing from those inferred from emission line modeling. Finally, we discuss the application of this technique to NASA's mission concept EVE, with the goal to provide simultaneous measurements of inner disk truncation radii, corotation radii and mass accretion rates for hundreds of young stars across the Galaxy. The unprecedented survey of inner disk properties that the mission would produce would enable the first stringent test
of angular momentum evolution theories in young stars and reveal the impact of the inner disk conditions on early planet architectures.

\end{abstract}


\section{Introduction} \label{sec:intro}

Mass accretion from a circumstellar disk represents a key stage in the early evolution of young, low-mass stars \citep[T Tauri stars;][]{hartmann2016}. This process leaves a direct and long-lasting imprint on defining stellar properties like mass and angular momentum, and it shapes the physical conditions in the inner disk at the epoch of planet formation and migration. 

While early models of disk accretion in T Tauri stars assumed a Keplerian disk that extends down to the surface of the star and transfers material to the latter via a boundary layer \citep[e.g.,][]{bertout1988}, inconsistencies with the observed properties of accreting young stellar objects (YSOs) prompted a paradigm shift in favor of magnetospheric accretion \citep{bouvier2007}. In this scenario, the inner disk is truncated at a characteristic radial distance from the star where the ram pressure of the disk material is overcome by the pressure of the stellar magnetic field (with typical dipole strengths between $\sim$0.1--1~kG; \citealp{johnstone2014}). The flow of accreting material from the inner disk to the star is then regulated by the stellar field lines, leading to the creation of accretion columns or extended accretion curtains that impact the stellar surface at near free-fall velocity \citep[e.g.,][]{romanova2003}. This picture can successfully explain the distinctive features of a classical T Tauri star (CTTS; T Tauri star undergoing mass accretion), including the strong emission lines with redshifted absorption components (inverse P~Cygni profiles; e.g., \citealp{edwards1994,reipurth1996}), the excess continuum emission observed from ultraviolet (UV) to infrared (IR) wavelengths \citep[e.g.,][]{gullbring2000,fischer2011}, and the pronounced photometric and spectroscopic variability \citep[e.g.,][]{fischer2023}.

In the magnetospheric accretion paradigm, the inner disk truncation radius, $R_T$, is a critical parameter to predict the evolution of star-disk interaction. In particular, the pattern of mass accretion is determined by the respective locations of $R_T$ and the inner disk corotation radius, $R_{CO}$, which corresponds to the orbital distance where the Keplerian angular velocity of the disk equals the stellar rotation rate. Competing angular momentum transfer mechanisms balance torque to determine the relative positions of these two key spatial scales \citep{ireland:2022}. If $R_T > R_{CO}$, the stellar magnetosphere rotates at a faster rate than the Keplerian angular velocity at the inner disk edge, and the resulting torque acting on the slower inner disk material prevents steady accretion from the disk to the star. In this configuration, the star-disk system is in a propeller regime \citep{ustyugova2006}, characterized by strong time variations in the accretion efficiency \citep{zhu2025}: disk material cyclically piles up at the magnetospheric radius until the gravitational acceleration prevails over the magnetospheric centrifugal barrier and the accumulated material is released in an episodic burst of accretion onto the star \citep{lii2014}. If $R_T \lesssim R_{CO}$, stable funnel-flow accretion may develop, with two large-scale magnetospheric accretion streams \citep[e.g.,][]{romanova2004}. However, if $R_T \ll R_{CO}$, i.e. when ram pressure greatly exceeds magnetic pressure, the inner disk edge is predicted to be unstable, leading to more stochastic accretion \citep{kulkarni2008} with multiple streams that can be transient in nature (chaotic unstable) or merge into one or two persistent streams (ordered unstable; \citealp{blinova2016}). More generally, accurate estimates of $R_T$ are key to constraining the mass accretion rate onto the star (\macc{}; \citealp{gullbring1998}), which in turn provides an essential diagnostic of mass transport processes through the disk \citep{alexander2023}, and the evolution of $R_T$ is expected to affect the orbital distribution of planets in the inner astronomical unit (AU) around the host star \citep{liu2017,romanova2019,mendigutia2024}.

In spite of its importance, the $R_T$ parameter is considerably more difficult to measure than $R_{CO}$ (the latter only requiring a photometric rotation period and stellar mass estimate). The small spatial scales associated with the magnetospheric accretion geometry around CTTSs prevent a direct determination of $R_T$ for all except the closest YSOs, for which near-IR (NIR) interferometry can help resolve the inner sub-AU disk emitting regions \citep[e.g.][]{eisner2014,bouvier2020,gravity2020}. Model-based estimates of $R_T$ include fitting emission lines profiles like H$\alpha$ with magnetospheric accretion flow models \citep{Thanathibodee2023,pittman2025}, although their accuracy can be limited by optically thick lines, which are common for typical accretion rates \citep{muzerolle2001}. Theoretical formulations of $R_T$ depend on properties that are not usually directly measurable \citep[e.g.,][]{bessolaz2008,pantolmos2026}, in particular the stellar magnetic field strength and configuration (detailed reconstructions of which are only available for representative objects in nearby regions), and the accretion rate in the disk. Often, a typical value $R_T \sim 5\,R_\star$ is applied uniformly across entire populations of low-mass, young stars, but this reference value does not reflect the diversity of young star–disk systems, including potential mass-dependent trends in the magnetic field topology \citep[e.g.,][]{gregory2012}.

In this paper, we present and test a new photometric diagnostic for $R_T$ around accreting young stars. Our method builds upon the work of \citet{robinson2021}, who combined 1D hydrodynamic simulations of the accretion columns, 1D plane-parallel, three-component accretion shock models, and a geometric rotational modulation model to create synthetic light curves of CTTSs and explore their morphology dependence on different input parameters. These models predict relations between the $R_T$ value and the colors of accretion-dominated sources on various photometric color-magnitude diagrams (CMDs). Because the separation between the inner disk edge and the stellar surface dictates the magnetospheric path followed by the accretion stream and the resulting kinetic energy of the accretion flow, $R_T$ has a crucial effect on the distribution of accretion shocks at the stellar surface \citep[e.g.,][]{romanova2011}, the properties of which can be discriminated by examining how the stellar colors change over a range of photometric brightness \citep[e.g.,][]{venuti2015}. 

To test these model predictions, we used the Las Cumbres Observatory Global Telescope network (LCOGT) in tandem with the Transiting Exoplanet Survey Satellite (TESS; \citealp{ricker2015}) to photometrically monitor stars in the Taurus and Upper Scorpius associations (Section~\ref{sec:data}). Thanks to their youth and proximity, these complexes have long been a benchmark for star formation and accretion studies (see, e.g., \citealp{kenyon2008} and \citealp{Preibisch2008}, respectively, for a review). We map the color variations of LCOGT sources on rotational timescales, using the TESS data to assess the fractional flux coverage of LCOGT observations; we then fit the observed color trends with the \citet{robinson2021} models and extract the best-fit parameters (Section~\ref{sec:models}). Section~\ref{sec:results} presents our model validation on YSOs with available $R_T$ estimates from interferometry and an analysis of our Taurus/Upper Scorpius pilot program in the context of pre-existing $R_T$ surveys. Finally, in Section~\ref{sec:eve}, we discuss future applications of this technique to accomplish one of the core scientific goals of NASA's EVE mission concept.

\section{Observations and data processing} \label{sec:data}

\subsection{Stellar sample} \label{sec:sample}

We have selected 19 disk-bearing YSOs in the $\sim$2~Myr-old Taurus star-forming region \citep{luhman2023} and seven in the $\sim$10~Myr-old Upper Scorpius \citep{luhman2025}, of which we conducted multi-band monitoring at LCOGT in $u',g',r',i'$ (henceforth $u,g,r,i$) bands ($\sim$350-770~nm), simultaneously with TESS high-cadence photometric monitoring. 

Kepler/K2 \citep{howell2014} observed over 200 young stars in Taurus during campaigns 4 and 13 (7~February to 23 April 2015 and 8 March to 27 May 2017, respectively). \citet{cody2022} extracted light curves for 155 confirmed Taurus members from the \citet{luhman2023} census, including 94 with IR excesses symptomatic of disks \citep{rebull2020}. They performed a detailed classification of K2 light curve profiles for the disk-bearing YSO sample and implemented the statistical metrics of \citet{cody2014} to assign each YSO to a distinct variability category, from bursting and stochastic (presumably driven by an erratic accretion pattern; \citealp{stauffer2014,stauffer2016}), to periodic and quasi-periodic (linked to long-lived distributions of magnetic and accretion spots at the stellar surface), to dipping (dominated by partial occultations of the stellar surface by vertical structures in the inner disk; \citealp{mcginnis2015}). 

We used \citeauthor{cody2022}'s (\citeyear{cody2022}) results to build an initial list of suitable Taurus targets with dynamics governed by mass accretion and star-disk interaction, removing sources with variability dominated by dust occultation signatures \citep{roggero2021}, as defined by a quasi-periodic or aperiodic dipper classification. We further removed all YSOs with physical or chance companions within 1 pixel (21$\arcsec$) radius and a 2~mag brightness range, which would appear as blends on the TESS images, or with Gaia re-normalized unit weight error (RUWE) statistic larger than 2.5 \citep{fitton2022}. To optimize the LCOGT monitoring campaign, we excluded all sources with $u$-band magnitudes from the literature (e.g., the most recent data releases of the Sloan Digital Sky Survey available at the time of the observations, namely SDSS DR16, \citealp{ahumada2020}, and SDSS DR18, \citealp{almeida2023}) that would preclude 10\% photometric precision with integration times $\leq$ 30 minutes. We also favored more spatially clustered targets to minimize the number of required LCOGT fields.   

Similar criteria were adopted to select suitable target stars in the Upper Scorpius association. Previous K2 monitoring data of this region \citep{cody2018} were used as reference to select disk-bearing members with variability dominated by accretion signatures (quasi-periodic, bursting or stochastic). We excluded all sources fainter than 15 in the $i$-band or for which a 10\% precision in the $u$-band would be unattainable at the LCOGT with integration times $\leq$30 minutes (as estimated using the LCOGT Exposure Time Calculator).

The final list of observed targets is reported in Table~\ref{tab:sample}. The vast majority of them have spectral types M0-M6 and typical distances between 130-160 pc. Most of the targeted objects have mid-infrared colors that categorize them as Class~II sources, while some are either more embedded (Class~I) or more evolved (Class II/III). The stars display mainly periodic or quasi-periodic variability in their TESS light curves (Sect.~\ref{sec:tess}); about 30\% have irregular, accretion-dominated variability (stochastic, bursting). Mass estimates for all sources are reported in Table~\ref{tab:log}. Unless noted otherwise, these were derived by fitting the sources' effective temperatures $T_\mathrm{eff}$ and bolometric luminosities $L_\mathrm{bol}$ on the H-R diagram with the pre-main sequence (PMS) MESA Isochrones \& Stellar Tracks (MIST; \citealp{dotter2016,choi2016}). $T_\mathrm{eff}$ values were assigned to each source based on their spectral type from \citet{luhman2023} for Taurus YSOs and \citet{luhman2025} for Upper Scorpius YSOs, using the conversion scale by \citet{gray2009}. $L_\mathrm{bol}$ values were derived from $J$-band magnitudes, using the NIR extinction estimates provided by \citet[][Taurus]{esplin2019} and \citet[][Upper Scorpius]{luhman2020}, and the PMS bolometric correction scale as a function of $T_\mathrm{eff}$ reported by \citet{pecaut2013}.

\begin{deluxetable*}{c c c c c c c c c}
\tablecaption{List of disk-bearing objects investigated for this study, grouped by parent region, with their TESS, 2MASS, and common identifiers, coordinates, distance, disk and light curve classifications, and spectral types.\label{tab:sample}}
\tablehead{
\colhead{TIC} & \colhead{2MASS} & \colhead{Name} & \colhead{R.A.} & \colhead{Dec.} & \colhead{Dist.\tablenotemark{a}} & \colhead{Disk class\tablenotemark{b}} & \colhead{LC\tablenotemark{c}} & \colhead{SpT\tablenotemark{d}} \\
& & & [deg] & [deg] & [pc] & & & }
\startdata
\multicolumn{9}{l}{Targets in Taurus:}\\
\rule{0pt}{3ex}56519506 & J04124068+2438157 & & 63.16962 & 24.63762 & $147.4 \pm 0.5$ & II & P & M3.5 \\
456944380 & J04311578+1820072 & & 67.81583 & 18.33522 & $141.2 \pm 0.5$ & II/III & QPS? & M4.25 \\
353752582 & J04313747+1812244 & V1213 Tau &  67.90627 & 18.20677 & & II & QPS & M0 \\
353752575 & J04313843+1813576 & HL Tau & 67.91046 & 18.23274 & & I & S? & K4 \\
430326905 & J04315779+1821380 & V710 Tau A & 67.99085 & 18.36046 & $145.2 \pm 0.4$ & II & QPS & M3.3 \\
430326907 & J04315968+1821305 & LkHa 267 & 67.99875 & 18.35836 & $150 \pm 2$ & II & QPS & M2.5 \\
268324608 & J04321540+2428597 & Haro 6-13E  & 68.06427 & 24.48312 & $127.7 \pm 1.6$ & II & B/S? & M0 \\
397287296 & J04321606+1812464 & & 68.06700 & 18.21281 & $146.9 \pm 1.0$ & II & P & M6 \\
397287185 & J04322627+1827521 & & 68.10954 & 18.46441 & $146.4 \pm 0.7$ & II/III & P & M5.25 \\
397350897 & J04324107+1809239 & & 68.17122 & 18.15657 & $142.5 \pm 0.5$ & II & QPS & M5 \\
61259576 & J04334871+1810099 & DM Tau &  68.45311 & 18.16936 & $143.1 \pm 0.5$ & II/III & B & M3 \\
61259650 & J04335283+1803166 & & 68.47027 & 18.05451 & $144.2 \pm 0.7$ & II/III & QPS & M5 \\
245862464 & J04343128+1722201 & & 68.63043 & 17.37217 & $144.5 \pm 0.8$ & II/III & QPS & M4.25 \\
150002984 & J04382858+2610494 & DO Tau & 69.61915 & 26.18031 & $138.4 \pm 0.7$ & II & S & M0.3 \\
245907047 & J04384502+1737433 & & 69.68765 & 17.62865 & $146.5 \pm 0.6$ & II & B & M4.25 \\
150173438 & J04422101+2520343 & & 70.58761 & 25.34278 & $139.6 \pm 0.8$ & II & P/QPS & M4.75 \\
125914505 & J04423769+2515374 & DP Tau & 70.65708 & 25.26027 & & II & S & M0.8 \\
436614017 & J04465305+1700001 & DQ Tau &  71.72110 & 16.99998 & $195.1 \pm 0.6$ & II & B & M0.6 \\
436614005 & J04470620+1658428 & DR Tau &  71.77592 & 16.97850 & $191.8 \pm 1.3$ & II & B & K6 \\
\multicolumn{9}{l}{}\\
\multicolumn{9}{l}{Targets in Upper Sco/$\rho$ Oph:}\\
\rule{0pt}{3ex}280156884 & J15573430-2321123 & V1148 Sco & 239.39296 & -23.35342 & $148.3 \pm 0.7$ & II/III & QPS & M1 \\
12621616 & J15581270-2328364 & & 239.55296 & -23.47681 & $145.0 \pm 0.4$ & II/III & P/QPS & G2 \\
9596871 & J16012652-2301343 & & 240.36050 & -23.02628 & $141.9 \pm 0.9$ & II & P & M4.5\\
49039044 & J16095933-1800090 & & 242.49717 & -18.00256 & $135.3 \pm 0.7$ & II & P & M5 \\
49165221 & J16111534-1757214 & & 242.81392 & -17.95594 & $135.3 \pm 0.3$ & II & P/QPS & M1\\
203822330 & J16251690-2322030 & & 246.32037 & -23.36764 & $138.5 \pm 0.3$ & II/III & QPS & M0\\
203817808 & J16254289-2325260 & & 246.42862 & -23.42392 & $137.8 \pm 0.4$ & II & QPS & M2.25\\
\enddata
\tablenotetext{a}{Geometric distances derived by \citet{bailer-jones2021}, when available, with corresponding 1\,$\sigma$ uncertainties.}
\tablenotetext{b}{The classification of each source as Class~I (protostellar stage), Class~II (PMS stage), or Class~II/III (transitional disk stage) was derived based on its mid-IR colors from the AllWISE catalog \citep{cutri2013}, following the scheme presented in \citet{koenig2014} to map these colors to the underlying morphology of the stellar spectral energy distribution (SED), which is indicative of the disk evolutionary stage \citep{lada1987,greene1994}.}
\tablenotetext{c}{TESS light curve morphology: periodic (``P''), quasi-periodic symmetric (``QPS''), burster (``B''), or stochastic (``S'').}
\tablenotetext{d}{Adopted spectral type for each source, extracted from \citet{luhman2023} for Taurus and \citet{luhman2025} for Upper Scorpius, which the exception of TIC 203822330 \citep{rizzuto2015} and TIC 203817808 \citep{esplin2018}.}
\end{deluxetable*}

\subsection{TESS} \label{sec:tess}

Taurus targets were observed with TESS throughout Sectors 43-44 (16 September to 6 November 2021) and/or 70-71 (20 September to 11 November 2023). Upper Scorpius targets were monitored with TESS during Sector~91 (9 April to 7 May 2025). Light curves for our targets were extracted from the TESS full frame images (FFI), which were taken at a cadence of 10 minutes through Cycle~4 (including Sectors 43-44), and 200 seconds from Cycle~5 (including Sectors 70-71 and 91). We used the \texttt{lightkurve} package to download $15\times$15-pixel cut-outs of the FFIs centered on each star of interest. We created a custom mask encompassing all points associated with the star for which the flux values exceeded eight standard deviations above the median level. If this mask included a close neighbor object, we then reduced its size to a $2\times$2-pixel box, again centered on the star of interest.  We used standard \texttt{lightkurve} background determination procedures, excluding pixel in the central star and any neighbors. Upon subtracting this background from the flux in the aperture, we further removed systematic trends by employing \texttt{lightkurve}'s RegressionCorrector function to determine the first two principle component vectors present in the background pixels. These were then subtracted from the flux time series to create the final light curves. 

\subsection{LCOGT} \label{sec:lco}

The multi-band LCOGT observations used for this work were acquired under three separate programs, each simultaneous with one of the TESS epochs discussed in Sect.~\ref{sec:tess}: program NSF2021B-020 (PI: L.~Venuti), program LCO2023B-002 (PI: A.M.~Cody), and program LCO2025A-008 (PI: A.M.~Cody). Each program used the Sinistro imaging cameras available on the LCOGT 1-meter telescope network. Individual observing blocks consisted of a set of $u,g,r,i$ exposures, and they were automatically assigned to the best telescope available by the LCOGT scheduling software, subject to the specified constraints on airmass (limiting value of 1.3), Moon-target angular separation ($>$30$^\circ$), and allowable observing windows (to ensure adequate spacing between repeated observations). LCOGT sites used for our programs include the Teide Observatory, the McDonald Observatory, the Siding Spring Observatory, and the South African Astronomical Observatory.

Details on the observations that were acquired at LCOGT for each target are reported in Table~\ref{tab:log}. While the two Taurus programs had distinct target lists and observing strategies, the observing fields were partially overlapping, which resulted in 40\% of targets having monitoring data from both semesters (2021B and 2023B). The 2021B observing program had the primary goal to reconstruct the changes in color exhibited, over rotational timescales, by disk-bearing YSOs with quasi-periodic flux behaviors over rotational timescales. Consequently, observations were scheduled with a daily cadence for ten consecutive days, in order to span the typical duration of 1–3 full rotational cycles for these young sources \citep{rebull2020}. Conversely, the 2023B program targeted sources with erratic flux behaviors due to episodic accretion, and the observing strategy was designed so as to map the entire color variation trends over the characteristic accretion burst recurrence timescales for each target \citep{cody2022}, with a cadence short enough to reconstruct the full photometric evolution of the brightening event produced by a discrete accretion episode. These distinctions explain the difference in number of visits accumulated for any overlapping source between 2021B and 2023B. The observing strategy for the 2025A Upper Scorpius program mirrored the one implemented for the 2023B Taurus program. Typical light curve durations range between 7.9--10.1 d for data acquired in 2021B, 15.3--53.9 d in 2023B, and 9.3--31.1 d in 2025A, with typical observing cadences in the range of 0.99--1.03~d, 0.75--2.03~d, and 0.13--0.32~d, respectively.

\begin{deluxetable*}{c c c c c c c c}
\tablecaption{Mass estimates, LCOGT observation log (observing semester, range of observing dates, number of photometric epochs per filter), fractional brightness phase coverage relative to the TESS light curves, color combinations used for the $R_T$ fit, and derived truncation radius estimates for targets in Table~\ref{tab:sample}.}
\label{tab:log}
\tablehead{
\colhead{TIC} & \colhead{$M_*$} & \colhead{Sem.} & \colhead{Dates (start/end)} & \colhead{$u/g/r/i$} & \colhead{Flux range} & \colhead{Colors used} & \colhead{$R_T$*} \\
 & [$M_\odot$] & & & &  & & [$R_\star$]}
\startdata
56519506 & $0.23_{-0.04}^{+0.05}$ & 2021B & 2021-09-16/2021-10-02 & 3/10/9/10 & 0.04--0.98 & $g-r/i$ & $3.2^{+0.1}_\mathit{-0.2}$\\
456944380 & $0.22_{-0.04}^{+0.04}$ & 2021B & 2021-09-30/2021-10-10 & 10/10/10/7 & 0.16--0.92 & $u-r/i$, $g-r/i$ & --\\
 & & 2023B & 2023-09-24/2023-11-17 & 45/45/45/47 & 0.01--1.00 & $u-r/i$, $g-r/i$ & --\\
353752582 & $0.45^{+0.14}_{-0.14}$\tablenotemark{\textnormal{\dag}} & 2021B & 2021-09-30/2021-10-10 & 9/9/10/7 & 0.09-0.97 & $u-r/i$, $g-r/i$ & $6.6^{+0.4}_{-0.2}$\\
 & & 2023B\tablenotemark{a} & 2023-10-14/2023-11-16 & 26/3/4/48 & 0.02--0.96 & $u-i$ & -- \\
353752575 & $1.11_{-0.11}^{+0.10}$ & 2021B & 2021-09-30/2021-10-10 & 10/10/10/5 & 0.02-0.92 & $u-r$, $g-r$ & $3.6^{+0.1}_{-0.1}$ \\
 & & 2023B & 2023-09-24/2023-11-17 & 70/58/54/38 & 0.00--0.96 & $u-r/i$, $g-r/i$ & --\\
430326905 & $0.270_{-0.007}^{+0.001}$ & 2021B & 2021-09-30/2021-10-10 & 10/10/10/7 & 0.00-0.87 & $u-r/i$, $g-r/i$ & $3.2^{+0.1}_{-0.1}$\\
 & & 2023B & 2023-09-24/2023-11-17 & 71/45/46/45 & -- & -- & --\\
430326907 & $0.32_{-0.04}^{+0.04}$ & 2021B & 2021-09-30/2021-10-10 & 6/7/10/7 & 0.03--0.84 & $u-r/i$, $g-r/i$ & $5.8^{+0.4}_{-0.2}$ \\
 & & 2023B\tablenotemark{b} & 2023-10-10/2023-11-17 & 35/2/7/40 &  0.11--0.92 & $u-i$ & $6.0^{+0.1}_{-0.1}$\\
268324608 & $0.47_{-0.05}^{+0.07}$ & 2021B & 2021-10-06/2021-10-16 & 7/9/9/10 & 0.01--0.71 & $u-r/i$, $g-r/i$& $5.8^{+0.1}_{-0.1}$ \\
397287296 & $0.113_{-0.013}^{+0.012}$ & 2021B & 2021-09-30/2021-10-10 & 6/10/10/7 & 0.17-1.00 & $u-r/i$, $g-r/i$& $\mathit{3.0^{+0.2}}$\\
397287185 & $0.13_{-0.02}^{+0.02}$ & 2023B & 2023-09-24/2023-11-17 & 14/41/44/46 & 0.00--0.96 & $u-r/i$, $g-r/i$ & --\\
397350897 & $0.13_{-0.01}^{+0.03}$ & 2023B & 2023-09-24/2023-11-17 & 18/41/42/42 & 0.02-0.97 & $u-r/i$, $g-r/i$ & --\\
61259576\tablenotemark{c} & $0.31_{-0.08}^{+0.02}$ & 2021B & 2021-09-30/2021-10-11 & 10/10/10/10 & 0.05--0.82 & $u-r/i$, $g-r/i$ & --\\
 & & 2023B & 2023-10-26/2023-11-07 & 15/6/5/5 & 0.17--0.79 & $u-r/i$, $g-r/i$ & --\\
61259650 & $0.14_{-0.02}^{+0.02}$ & 2021B & 2021-09-30/2021-10-11 & 2/10/10/10 & 0.11--0.84 & $g-r/i$ & $\mathit{3.0^{+0.2}}$\\
245862464\tablenotemark{d} & $0.22_{-0.04}^{+0.05}$ & 2023B & 2023-09-25/2023-11-16 & 14/15/16/16 & 0.00--0.96 & $u-r/i$, $g-r/i$ & $3.2^{+0.1}_\mathit{-0.2}$\\
150002984\tablenotemark{e} & $0.39_{-0.02}^{+0.06}$ & 2021B & 2021-10-12/2021-10-21 & 3/10/10/4 & 0.20--0.93 & $g-r$ & $7.6^{+0.8}_{-0.4}$\\
245907047 & $0.22_{-0.04}^{+0.05}$ & 2021B & 2021-10-12/2021-10-21 & 9/10/9/10 & 0.28-0.90 & $u-r/i$, $g-r/i$ & --\\
150173438 & $0.14_{-0.01}^{+0.03}$ & 2021B & 2021-10-27/2021-11-05 & 4/9/9/9 & 0.11-0.74 & $g-r/i$ & $3.2^{+0.1}_\mathit{-0.2}$ \\
125914505 & $0.47_{-0.06}^{+0.07}$ & 2021B & 2021-10-27/2021-11-05 & 4/9/9/9 & 0.03-0.84 & $u-r/i$, $g-r/i$ & $4.8^{+0.1}_{-0.2}$ \\
436614017 & $0.40_{-0.03}^{+0.03}$ & 2021B & 2021-10-12/2021-10-21 & 10/10/10/10 & 0.00-0.95 & $u-r/i$, $g-r/i$ & $3.6^{+0.1}_{-0.2}$ \\
 & & 2023B & 2023-10-11/2023-10-27 & 16/16/15/15 & 0.00--0.93 & $u-r/i$, $g-r/i$ & $3.4^{+0.2}_{-0.1}$\\
436614005 & $0.77_{-0.16}^{+0.16}$ & 2021B & 2021-10-12/2021-10-21 & 10/10/10/10 & 0.07--0.97 & $u-r/i$, $g-r/i$ & $6.8^{+1.2}_{-0.2}$\\
 & & 2023B\tablenotemark{f} & 2023-10-11/2023-10-27 & 16/16/15/15 & 0.02-0.92 & $u-r/i$, $g-r/i$  & $5.2^{+0.4}_{-0.1}$\\
280156884 & $0.40_{-0.04}^{+0.06}$ & 2025A & 2025-04-18/2025-05-09 & 70/74/53/57 & 0.01-0.97 & $u-r/i$, $g-r/i$ & --\\
12621616 & $1.410_{-0.014}^{+0.014}$ & 2025A & 2025-04-18/2025-05-09 & 73/26/10/16 & 0.04-0.96 & $u-r/i$, $g-r/i$  & --\\
9596871 & $0.17_{-0.03}^{+0.04}$ & 2025A & 2025-04-10/2025-04-20 & 0/27/27/27 & 0.03-0.83 & $g-r/i$ & $\mathit{3.0^{+0.2}}$\\
49039044 & $0.126_{-0.011}^{+0.032}$ & 2025A & 2025-04-29/2025-05-08 & 1/29/30/30 & 0.06-0.74 & $g-r/i$ & $\mathit{3.0^{+0.1}}$\\
49165221 & $0.42_{-0.04}^{+0.06}$ & 2025A & 2025-04-29/2025-05-08 & 32/29/31/30 & 0.05-0.98 & $u-r/i$, $g-r/i$ & --\\
203822330 & $0.47_{-0.04}^{+0.08}$ & 2025A & 2025-04-10/2025-05-11 & 26/92/91/92 & 0.11--0.99 & $u-r/i$, $g-r/i$ & --\\
203817808 & $0.32_{-0.04}^{+0.05}$ & 2025A & 2025-04-10/2025-05-11 & 15/48/71/76 & 0.13--1.00 & $u-r/i$, $g-r/i$ & $7.8^{+0.2}_{-0.1}$\\
\enddata
\tablenotetext{\ast}{Italics indicate upper limits.}
\tablenotetext{\dag}{Estimate extracted from \citet{lopez_vazquez2024}.}
\tablenotetext{a}{The fit returns multiple possible solutions with similar probabilities between $R_T = 4.2\ R_\star$ and $R_T = 6.8\ R_\star$.}
\tablenotetext{b}{For this source and epoch, the marginalized probability shows a secondary peak at $R_T = 7.6^{+0.2}_{-0.4}$.}
\tablenotetext{c}{Although the fitting procedure returns a solution around $R_T \sim 4.2\ R_\star$ for this source, there is no clear probability peak above the noise level.}
\tablenotetext{d}{The probability map for this source also displays an increase in value at the higher end of the investigated $R_T$ domain, although without a clear peak.}
\tablenotetext{e}{For this source, an additional probability peak of similar intensity as the first peak appears around $R_T = 6.2^{+0.1}_{-0.4}\ R_\star$.}
\tablenotetext{f}{For this source and epoch, the marginalized probability shows a secondary peak at $R_T = 7.2^{+0.1}_{-0.4}$.}
\end{deluxetable*}

Point source catalogs for each exposure and filter were retrieved from the data products of the LCOGT data reduction pipeline, BANZAI \citep{banzai}, which encompasses steps of bias and dark subtraction, flat field correction, source extraction, and astrometric calibration. Targets of interest from the \citet{luhman2023} Taurus census were cross-matched with the LCOGT single-epoch catalogs by allowing a matching radius of 1$\arcsec$ on (R.A.,~Dec.), and any flux measurements associated with an estimated background $>$3\,$\sigma$ above the median background level on the corresponding frame were rejected. A similar approach was adopted for Upper Scorpius fields, using the \citet{luhman2025} census as a reference. To correct the instrumental light curves for nightly zeropoint offsets, we built a reference zeropoint curve from the median-subtracted light curves of all field stars detected in the same field of view as the target stars. Typically, four reference stars were used for any given field of view, the non-variable nature of which was verified by inspecting their TESS light curves, as released from the Quick-Look Pipeline (QLP; \citealp{huang2020, kunimoto2022}). The zeropoint-corrected instrumental light curves for target stars were then converted to apparent SDSS magnitudes by taking as photometric standard the reference star in the corresponding field that exhibited the least photometric scatter, across all filters, after subtraction of the reference zeropoint curve.

In a situation affecting the differential photometry derivation for TIC 245862464, 245907047, 245907051, 436614017, and 436614005, no $u,g,r,i$ photometric set could be retrieved from the SDSS database for any of the reference stars in the field. In those cases, we extracted $g,r,i$ magnitudes and associated uncertainties from the catalog assembled by \citet{huber2016}, and used the color relations tabulated by \citet{davenport2014} to estimate the $u$-band magnitude for each reference star from the observed $g$-band magnitude and the tabulated $u-g$ color corresponding to the ($g-i$, $g-r$, $r-i$) point on the main sequence color locus that lies closest to the observed colors, after correcting for extinction. To account for the observational uncertainties, 10\,000 sets of $g,r,i$ magnitudes were extracted at random from normal distributions centered around the best magnitude values and with widths equal to the associated uncertainties. For each simulated $g,r,i$ set, we determined which point, on the reference color sequence, best reproduced the simulated colors with a common extinction solution and single-band extinction coefficients calculated by the Spanish Virtual Observatory (SVO) Filter Profile Service \citep{svo}. We then derived the apparent $u$-band magnitude corresponding to the simulated $g,r,i$ magnitudes, and used the results from all simulated sets to build a statistical distribution of $u$-band values, the average and standard deviation of which were assumed as best $u$-band estimate and associated uncertainty for the reference star. 

A total of 45 Taurus YSOs were observed during our LCOGT 2021B+2023B campaigns: 19 that satisfy all of our criteria discussed in Sect.~\ref{sec:sample} and thus constitute the primary focus of this work; three that exhibit intrinsic and variability properties suitable for our investigation but have no suitable accompanying TESS data; eight that are unresolved binaries or with RUWE values above our adopted threshold for presumably single stars; eight that are known dipper-like variables; and seven with infrared classifications as Class~III sources or naked photospheres \citep{teixeira2012}. In Upper Sco, 42 YSOs were observed, 20 of which are classified as disk-bearing sources. Among these, 13 either lack reliable TESS light curves or display dipping variability, thus leaving seven sources that are suitable for our pilot study.\looseness=-1

\section{Models} \label{sec:models}

\subsection{Building the model tracks} \label{sec:model_description}

We constructed the color-magnitude model tracks using the simulations of \citet{robinson2017, robinson2021} with a few small modifications. We briefly describe the model and these modifications here. {A schematic of the model is shown in Fig.~\ref{fig:schematic}}. 
The simulations solve the hydrodynamic fluid equations along 1D curvilinear coordinates that follow along the magnetic field of the young star that truncates the inner disk. We assume the field is a rotation axis-aligned dipole. The resulting velocity and density of the flow reaching the star is used as input into a radiative transfer model of the shock region \citep{calvet1998, robinson2019} to produce the accretion excess. This model works by solving the density structure of the postshock region, which is then used to produce the downward and upward emission. This postshock emission, which is primarily X-rays, impinges on the the underlying photosphere and preshock region. Those regions reprocess the X-rays into UV and optical light, forming the observed accretion excess.  

This excess is then modulated as a function of rotational phase, following \citet{robinson2021}, by scaling the flux by the projected area of a strip with an azimuthal extent of $90^\circ$ and a polar extent set by assuming lower and upper bounds from a magnetic field line that connects to the disk at 0.5 and 1.5 times the reported value of $R_T$. 
We convolve the resulting excess and a synthetic photosphere with a filter passband to produce synthetic photometry. 

To eliminate shear from differential rotation at the truncation point and keep the number of simulations in our grid computationally tractable, we make the approximation that $R_T \approx R_{CO}$ while solving the fluid simulations. While we note that this is in conflict with the goal of measuring $R_T$ and $R_{CO}$, the only effect within the simulations is modifying the centrifugal force term within the fluid momentum equation. While this does have a small impact on the flow, the distribution of synthetic photometry points in color-color space of the excess is still largely determined by the truncation radius. We also note that other effects that arise from values of $R_T/R_{CO} \neq 1$ (e.g., magnetospheric ejections) are already ignored by modeling the system in 1D. 

The density boundary condition at the truncation radius is set by approximating the disk as a thin, non-isothermal Shakura-Sunyaev $\alpha$-disk \citep{shakura1973} emitting as a blackbody. This results in the following radial density profile
\begin{equation}
    \frac{\rho(R)}{\rho_0} = \frac{R_0^{15/8}}{R^{15/8}}\frac{(1 - (R_\star/R)^{1/2})^{5/8}}{(1 - (R_\star/R_0)^{1/2})^{5/8}}
\end{equation}
where $\rho_0$ is a reference density set at a reference radius of $R_0 = 5 \, R_\star$, chosen such that we achieve the approximate observed range of kinetic energy fluxes, $F = \frac{1}{2} \rho u^3$, for material striking the surface of the star across the model grid
\citep[$\log_{10}(F) \approx 9.5$ to $12$ for F in $\mathrm{erg \, s^{-1} \, cm^{-2}}$;][]{robinson2019, pittman2022}. Following \citet{robinson2021}, the inner velocity boundary condition is set to mimic Kolmogorov turbulence by using a velocity driving function. This function has a turbulent amplitude of 0.1 $c_s$, as the amplitude should approximately scale as $\sqrt\alpha$, whose value is typically inferred to be $\alpha \sim 0.01$ for disks \citep[e.g.,][]{hartmann1998}.

The grid of models used to fit the data includes stellar masses of $M_\star = 0.1,\, 0.5, \, 1.0,\, 1.5,$ and $2.0 \, M_\odot$, each evaluated at truncation radii of $R_T = 3.0$ to $9.0 \, R_\star$ in $0.2 \, R_\star$ steps. We construct light curves at inclinations $\theta$ between $0^\circ$ and $90^\circ$ in $10^\circ$ increments, resulting in a total of 1550 models. We estimated $R_\star$ and $T_\mathrm{eff}$ for each mass using the MIST isochrones at an age of 1 Myr at solar metallicity. We adopt the BT Settl models \citep{allard14} as synthetic non-accreting photospheres. We simulate the system for 30 days following a 15-day initialization period. We sample the simulation to produce synthetic photometry {in the SDSS $u,g,r,i$ passbands at a cadence of 30 minutes, resulting in 1440 photometry points per model. 

\begin{figure}
\centering
\includegraphics[width=0.47
\textwidth]{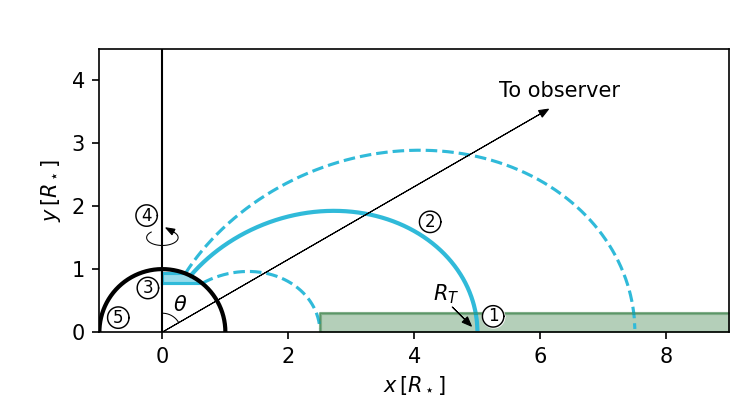}
\caption{Side-view of the model of magnetospheric accretion used to produce synthetic light curves. Each of the major components are labeled and described by the following models: 1) Non-isothermal \citet{shakura1973} $\alpha$-disk and velocity perturbations from Kolmogorov turbulence.  2) Velocity and density along the accretion funnel via the 1D-fluid simulations of \citet{robinson2017}. 3) Accretion excess from the shock model of \citet{calvet1998}. 4) Surface coverage and rotational modulation of the accretion hot spot following the prescription of \citet{robinson2021}. 5) Non-accreting photosphere from the BT Settl model grid \citep{allard14}.}
\label{fig:schematic}
\end{figure}

\begin{figure*}
\centering
\includegraphics[width=\textwidth]{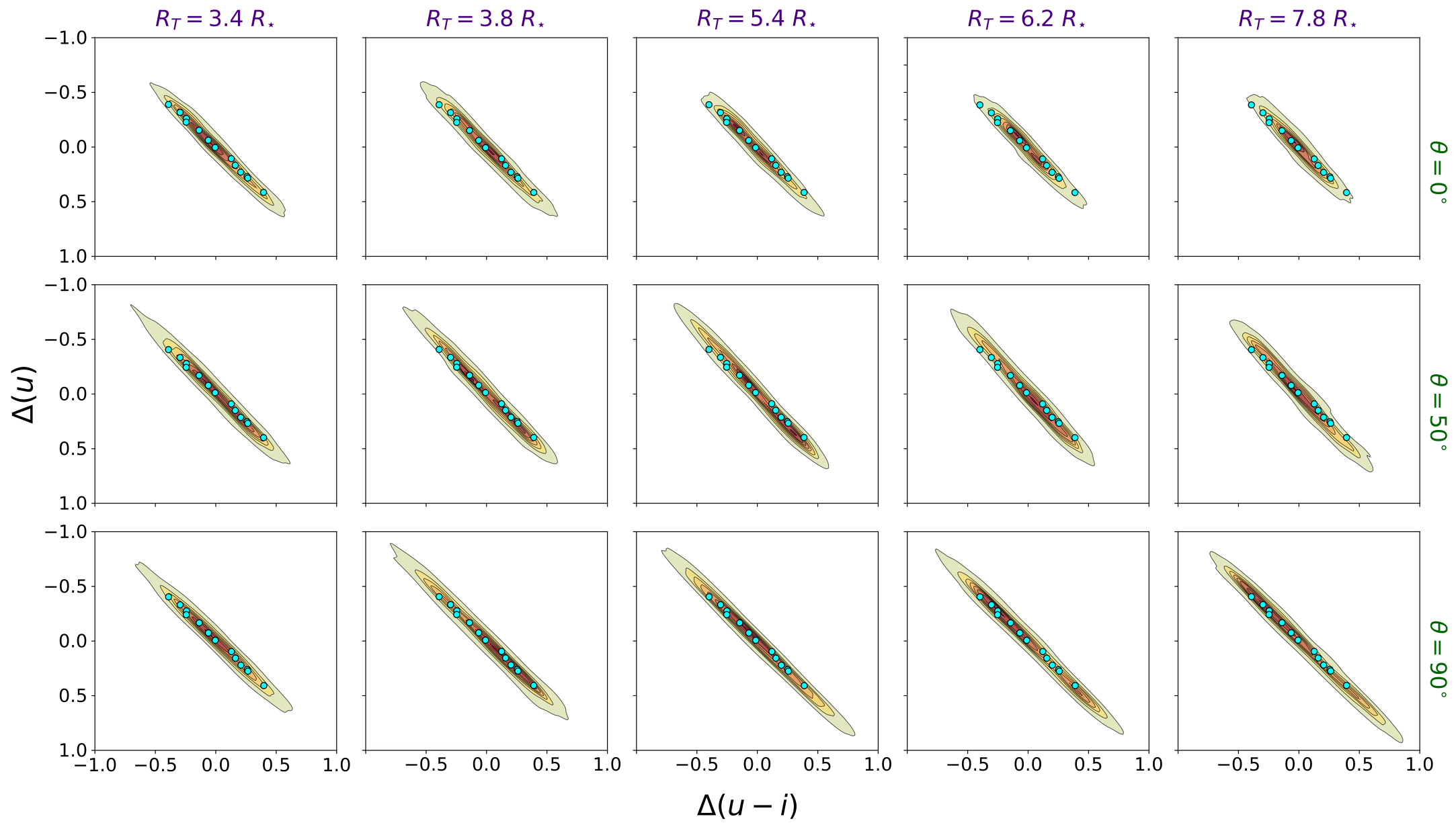}
\caption{Density contours extracted from the model tracks for a 0.5~$M_\odot$ star on the ($u-i, u$) CMD as a function of the assumed $R_T$ (left to right) and stellar inclination $\theta$ (top to bottom). The cyan dots correspond to the color properties of the disk-bearing star TIC~203817808, which we monitored at the LCOGT during semester 2025A (Table~\ref{tab:log}).The illustrated values of $R_T$ correspond to the local peaks in the probability distribution derived from the data-model fit as a function of $R_T$ (see Fig.~\ref{fig:Rt_fit_results}); the last column corresponds to the best-fitting value of $R_T$. Following the procedure described in Sect.~\ref{sec:model_fits}, the model contours have been broadened by an amount corresponding to the photometric uncertainty on the observations for TIC~203817808. Fits encompass a variety of model features, including the color slope, the variability amplitudes, and the datapoint distribution relative to the predicted areas of maximum density (dark brown contours).}
\label{fig:model_grid}
\end{figure*}

\subsection{Fitting models to data} \label{sec:model_fits}

To identify the model color tracks that best fit our LCOGT targets in Table~\ref{tab:sample}, we adopted the kernel density-based two-sample comparison test \texttt{kde.test} \citep{chacon2018}, implemented in the R package \texttt{ks} for multivariate kernel smoothing \citep{duong2007}. For each model track, we built a grid of density contours on a given CMD that reflects the probability of finding the observed source at a specific point on that CMD under the assumed stellar and magnetospheric parameters (Fig.~\ref{fig:model_grid}). This approach allows us to use all the distribution attributes (e.g., curvature, density, width) in the fit, providing a significant improvement over a simpler parametric fit. In order to account for the effect of observational uncertainties, prior to our data-model comparison we injected a random noise value into each simulated epoch of the model tracks, drawn from a normal distribution centered on 0 and with standard deviation equal to the observational 1~$\sigma$ uncertainty. 

For each target from Table~\ref{tab:sample}, we ran the fitting routine on the model tracks with input mass that lies closest to the $M_\star$ estimate listed in Table~\ref{tab:log}. We conducted the fit on four CMD planes ($m_b-m_r$, $m_b$), where $m_b$ is one of the bluer filters ($u,g$) and $m_r$ is one of the redder filters ($r,i$). Paired magnitude measurements in color indices are typically separated by an interval $\sim$1--3 minutes in $g-r/i$ and $\sim$3--15 minutes in $u-r/i$. Color and magnitude sequences were normalized to the time-series average prior to the fit. We used the simultaneous TESS light curves to estimate the fraction of the total flux variability amplitude covered by the LCOGT observations. This was achieved by assigning a brightness phase $\phi=\frac{F_\mathrm{TESS}(t)-\min(F_\mathrm{TESS})}{\max(F_\mathrm{TESS})-\min(F_\mathrm{TESS})}$ to each TESS flux measurement $F_\mathrm{TESS}(t)$ with epoch $t$ included in the LCOGT observation window. Each LCOGT epoch was then matched with the $\phi$ value of the closest TESS epoch. For each pair of filters, similar brightness phases were calculated for the theoretical time series, and only the portions of the simulated light curves that fell within the fractional flux range spanned by the observations were used to construct the theoretical CMD for the fit. This step helps prevent erroneous results where time series data that appear to vary less due to incomplete observational coverage of their flux excursions would be fit by models that predict small total variability amplitudes.

The KDE test returns the two-sided p-value corresponding to the calculated t-statistics between the two samples (data and model distribution). The best-fitting model is then the one for which the derived p-value is the largest (implying that the null hypothesis of observational and theoretical distributions being extracted from the same parent distribution is retained to a greater significance).
To account for fluctuations in p-value deriving from the injection of observational noise into the model tracks, and to extract an estimate of the uncertainty around the best-fit $R_T$, we repeated the fitting procedure 10 times and recorded the median ($\tilde{p}$), 16th percentile ($p_{16^{th}}$) and 84th percentile ($p_{84^{th}}$) in the resulting p-value distribution. To combine p-values from individual CMDs into a single statistic and extract the overall best-fitting $R_T$, we followed the \citet{stouffer1949american} z-score method (see also \citealp{heard2018}), using $\tilde{p}$ as the reference p-value for each model and filter combination, and weights $w$ defined by the logarithmic difference between the 84th and 16th percentile values ($w = [\log{(p_{84^{th}}/p_{16^{th}})}]^{-1}$). 

The procedure discussed above yields a 2D map of probability values over the theoretical ($R_T, \theta$) grid. Because our goal is to extract measurements of $R_T$, not $\theta$, we marginalized over a prior distribution on $\theta$, defined either as a Gaussian probability distribution centered on a given source's measured inclination from the literature, when available, or as an isotropic distribution of stellar inclinations. We weighed our initial probability map by the assumed $\theta$ distribution and then marginalized the resulting map over $\theta$ to obtain our final 1D $R_T$ diagnostic. We used a sigma-clipping procedure to calculate the noise floor and dispersion of the marginalized $R_T$ probabilities as their mean and standard deviation, respectively, and then extracted as significant all $R_T$ solutions above the noise+dispersion level.

\subsection{Theoretical fit precision}

\begin{figure*}
\centering
\includegraphics[width=\textwidth]{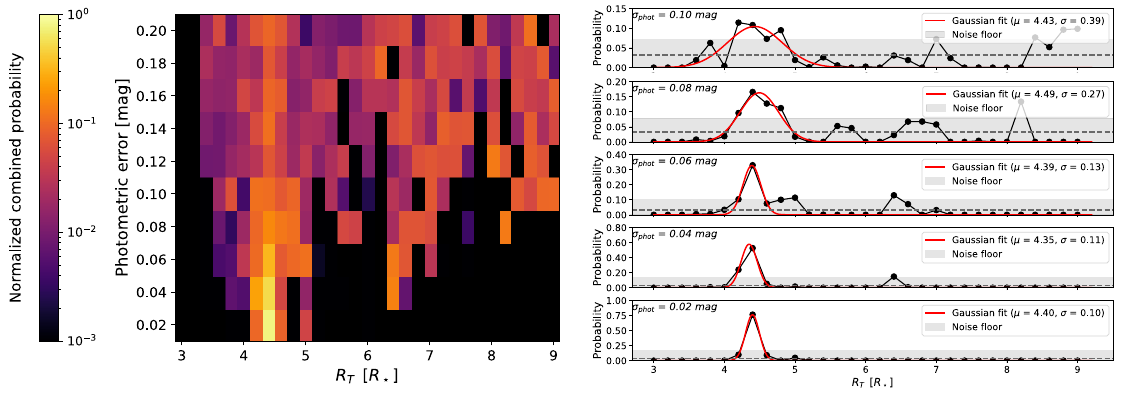}
\caption{\textit{Left}: Probability map that illustrates the recoverability of the true $R_T$ value with the model fits discussed in Sect.~\ref{sec:model_description} as a function of photometric uncertainty. A true value $R_T = 4.4\ R_\star$ was adopted for this test. Each row of the map shows the probability of the $R_T$ fit solution labeled on the $x$-axis, marginalized over stellar inclination, relative to the amount of simulated photometric noise that is labeled on the $y$-axis. The color scale is capped at probabilities $\geq 10^{-3}$ for clarity; the actual range extends down to $\sim 10^{-9}$. \textit{Right}: Detailed view of the ($R_T$, probability) sequences extracted from our sensitivity analysis for assumed photometric uncertainties of 0.02, 0.04, 0.06, 0.08, and 0.1 mag. The noise floor of the map is illustrated as a gray shaded area, while the dashed gray line marks the sigma-clipped average of the noise floor. The red curve shows, in each case, a Gaussian fit to the probability peak corresponding to the true value of $R_T$, with $\mu$ and $\sigma$ parameters identifying the best solution and its uncertainty.}
\label{fig:sensitivity_tests}
\end{figure*}

To test the impact of observational uncertainties on the sensitivity of our model fits for recovering the true $R_T$ values, we selected a model at random from our grid and generated synthetic observations by gradually injecting an increasing amount of photometric noise $\sigma_{phot}$ between 0.02--0.2 mag in steps of 0.02 mag. For each filter and epoch of the synthetic light curves, we randomly extracted a noise value from a Gaussian distribution centered on 0 with standard deviation $\sigma_{phot}$ and added it to the theoretical magnitude value. We then applied the procedure described in Sect.~\ref{sec:model_fits} to fit the simulated observations with the entire model grid at the corresponding stellar mass. We marginalized the derived probabilities over stellar inclination and conducted a Gaussian fit to the probability peak to determine whether the input $R_T$ is retrieved and to what precision.

Our results are illustrated in Fig.~\ref{fig:sensitivity_tests}. The reference model used to generate the synthetic data with noise assumes $M_\star = 0.5\ M_\odot$ and $R_T = 4.4\ R_\star$. The probability map (left panel) shows a well-localized, narrow probability peak around the true value for $R_T$ when the photometric uncertainty is $\sim$0.05 mag or less, and the corresponding Gaussian fits indicate that the true value is recovered to a precision $<$0.2~$R_\star$ (right panels). As the noise increases, the probability peak around $R_T \sim 4.4\ R_\star$ becomes wider and less prominent relative to the noise floor, with precisions $\sim$0.4~$R_\star$ on the recovered $R_T$ when the input photometric uncertainties amount to $\sim$0.1~mag. For photometric uncertainties $\sim$0.18--0.2~mag, the best $R_T$ solution identified via the model fits may not correspond to the true value.

\section{Results} \label{sec:results}

\subsection{Test cases: GRAVITY targets} \label{sec:gravity}

We validated our model performance against the results of the GRAVITY Guaranteed Time Observations (GTO) program dedicated to YSOs. GRAVITY \citep{gravity2017} is a second-generation instrument at the Very Large Telescope Interferometer that provides high spectral resolution in the $K$-band ($\sim$2.2~$\mu$m) with milliarcsecond angular resolution. These characteristics enable spatially resolved observations of Br$\gamma$ emission in CTTSs, largely produced by the accelerated gas in the magnetospheric accretion flow from the inner disk edge to the star (although spatially extended contributions associated with winds and outflows can also be detected; e.g., \citealp{beck2010}). Under the assumption that accretion dominates the Br$\gamma$ emission, these measurements provide a direct estimate of the size of the magnetospheric cavity that extends from the stellar surface to the truncated inner disk edge.

Among the other targets of the GTO YSO program, GRAVITY observed nine K- to early M-type CTTSs for which the size of the Br$\gamma$-emitting inner disk region could be resolved \citep{Wojtczak2023,Soulain2023,Perraut2026}. These include CI~Tau, in the Taurus region, which received LCOGT $g,r,i$ monitoring \citep{manick2024} that overlapped in time with the GRAVITY data \citep{Soulain2023}. We used these archival light curves to test our model inference against the GRAVITY Br$\gamma$ measurement for CI~Tau. The original light curves extend over three months, a timescale over which changes in $R_T$ can be triggered by changes in \macc{} \citep[e.g.,][]{armeni2024}. For consistency with the average duration of our targets' LCOGT light curves, we restricted our fit to a 20~d portion of the LCOGT data centered on the epoch of the GRAVITY observations. We adopted the same stellar parameters that were provided for CI~Tau by the GRAVITY collaboration, namely mass $M_\star = (0.9 \pm 0.02)\ M_\odot$ \citep{simon2019} and inclination $\theta = 71^\circ \pm 1^\circ$ \citep{Soulain2023}. Our results are illustrated in Fig.~\ref{fig:gravity_tests}. We detect a clear probability peak at $R_T = 4.8^{+0.2}_{-0.8} R_\star$, which matches the Br$\gamma$ half-flux radius of $4.8^{+0.8}_{-1.0} R_\star$ reported by \citet{Soulain2023}.

We also applied our fitting procedure to DoAr~44, another GRAVITY target, in the Ophiucus cloud, that was monitored at the LCOGT in $u,g,r,i$ filters at the same time as the GRAVITY observations \citep{bouvier2020phot}. Two epochs of GRAVITY data, taken on consecutive days, are available for this source, as reported by \citet{bouvier2020} who inferred upper limits of $5.0\,R_\star$ and $3.9\,R_\star$ for the Br$\gamma$ half-flux radius from the two individual epochs. The first of those two epochs was subsequently re-analyzed by \citet{Wojtczak2023}, who reported a Br$\gamma$ half-flux radius measurement of $(5.93 \pm 0.47) R_\star$. The LCOGT light curves for this source, the tail ends of which overlap with the GRAVITY observations, have similar duration ($\sim$20~days) and observing cadence (from $\sim$15 minutes to 1~day) as the data from our own LCOGT programs (Sect.~\ref{sec:lco}). Thus, we applied our model fitting routine, as described in Sect.~\ref{sec:model_fits}, to the original light curves retrieved from \citet{bouvier2020phot}, using the input parameters $M_\star = (1.5 \pm 0.2)\ M_\odot$ and $\theta = 32^\circ \pm 4^\circ$ \citep{Wojtczak2023}. From our analysis, we derive a measurement of $R_T = (3.8 \pm 0.1)\ R_\star$, consistent with the \citet{bouvier2020} results.

\begin{figure}
\centering
\includegraphics[width=0.47\textwidth]{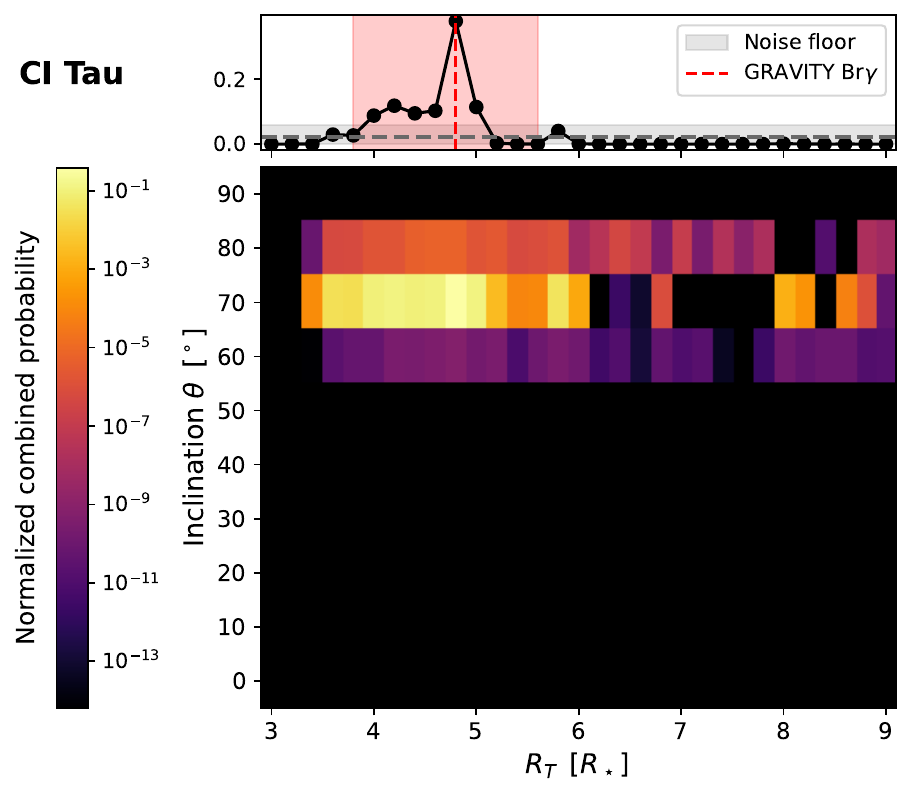}
\caption{Probability map of different $(R_T, \theta)$ solutions for CI~Tau across our model grid. The map shows combined probabilities from our four-CMD joint analysis as discussed in Sect.~\ref{sec:model_fits}, scaled by the probability distribution in $\theta$, and normalized so that the sum of probabilities across the map equals 1. The panel above the map shows the final probabilities marginalized over $\theta$; the gray shaded area marks the dispersion around the noise floor on the map, while the red dashed line and shaded area correspond to the GRAVITY Br$\gamma$ measurement and its uncertainty, respectively, as reported by \citet{Soulain2023}.}
\label{fig:gravity_tests}
\end{figure}

\subsection{Distribution of $R_T$}

\begin{figure*}
\centering
\includegraphics[width=\textwidth]{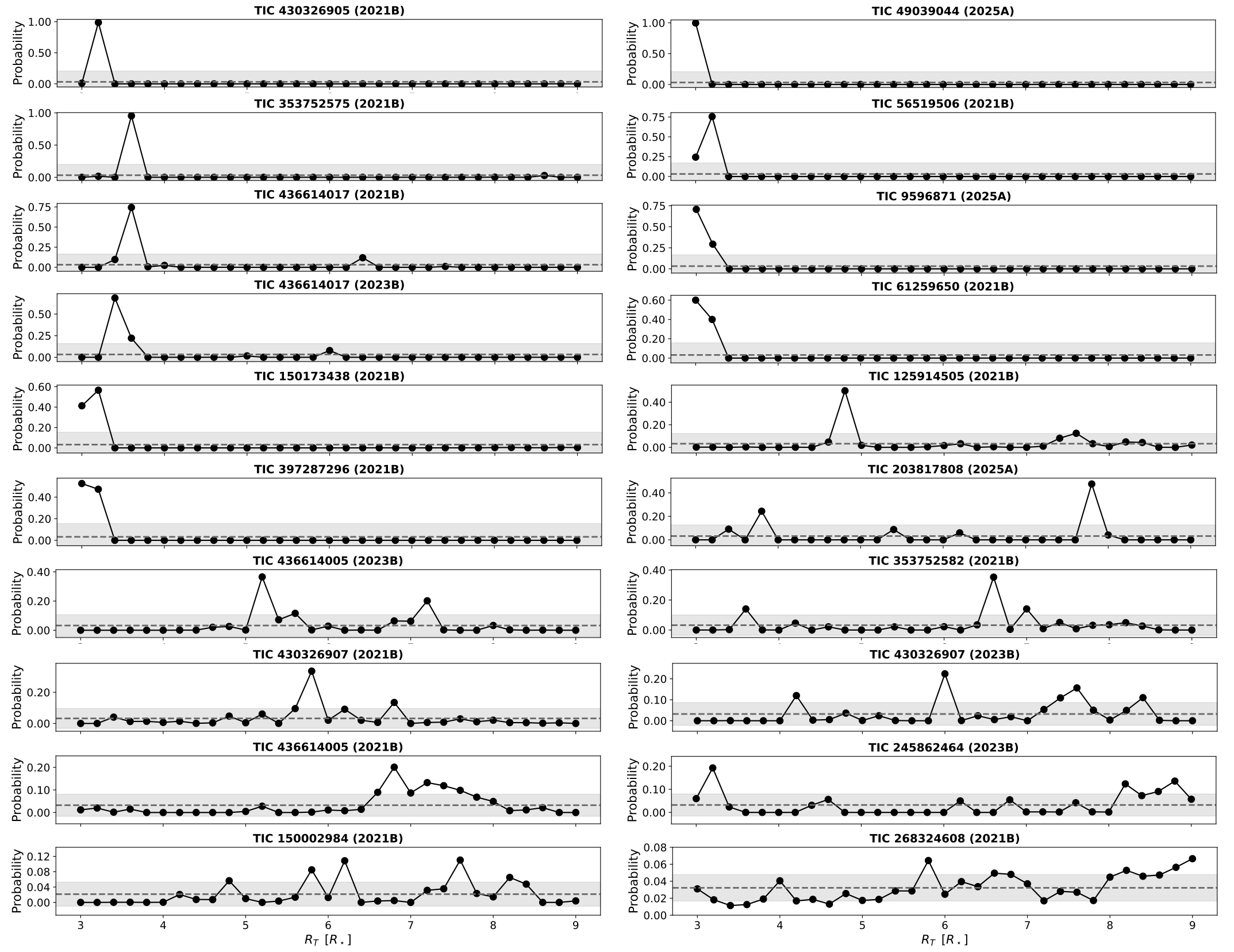}
\caption{Marginalized probabilities of different $R_T$ fit solutions for targets in Table~\ref{tab:log}. The order has been rearranged to illustrate the different morphologies from high-probability to low-probability solutions. On each panel, the gray shaded area corresponds to the noise level associated with the underlying probability map (see Sect.~\ref{sec:model_fits}).}
\label{fig:Rt_fit_results}
\end{figure*}

From our analysis, we derived an $R_T$ estimate for 17 of our 26 targets, and 20 of our 34 individual light curve sets (Table~\ref{tab:log}). The diagnostic diagrams used to infer these estimates are shown in Fig.~\ref{fig:Rt_fit_results}, ranked by probability at the peak. Of the nine sources with no $R_T$ fit solution (marginalized p-value $<$ 0.05), six are classified as more evolved (Class~II/III) in Table~\ref{tab:sample}, thus their accretion signatures may be less prominent than their photospheric signatures, leading to an unsuccessful or less reliable fit. Of the two Class~II/III sources in our list (Table~\ref{tab:sample}) for which an $R_T$ solution is provided in Table~\ref{tab:log}, one (TIC 61259650) has a best-fit $R_T$ that lies at the edge of the investigated domain ($R_T = 3.0\ R_\star$), and the other (TIC~245862464) has potential solutions at both ends of the investigated $R_T$ range.  

In spite of our limited sample, some trends emerge from our derived distribution of $R_T$ estimates. About half of our $R_T$ measurements (11) are concentrated in the $[3,3.6]\ R_\star$ range. Four sources have a best-fit $R_T = 3\ R_\star$, where our model grid ends, and we thus consider those estimates as upper limits. The remaining $R_T$ measurements are uniformly distributed in $R_T$ between 4.8--7.8~$R_\star$. The mass dependence of our $R_T$ estimates is shown in Fig.~\ref{fig:Rt_distribution}. All of the lowest mass stars in our sample ($M_\star \lesssim 0.3\ M_\odot$) appear in the pile-up of objects at small $R_T$, whereas no mass-dependent trend appears in the $R_T$ distribution for $M_\star > 0.3\ M_\odot$ targets. No obvious connection between $R_T$ values and variability classifications appear for the bulk of our sample, with the exception of stars labeled as periodic (``P''), which all correspond to $R_T$ estimates of 3--3.2~$R_\star$.

\subsection{Comparison with $R_T$ distributions from the literature}

\begin{figure}
\centering
\includegraphics[width=0.47\textwidth]{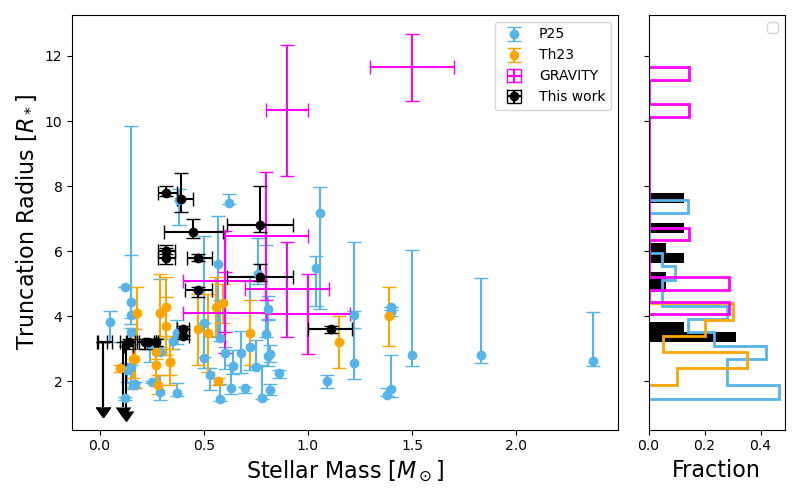}
\caption{Distribution of $R_T$ measurements derived in this work as a function of stellar mass, compared with results from \citet[]{Wojtczak2023}, \citet[][``Th23"]{Thanathibodee2023}, and \citet[][``P25"]{pittman2025}. Upper limits for several targets from this work are illustrated with arrows. The normalized $R_T$ distribution inferred from each study is illustrated in the side histogram.}
\label{fig:Rt_distribution}
\end{figure}

In this section, we examine our inferred distribution of $R_T$ values in the context of statistical results from previous surveys. We focus on the interferometric results from the GRAVITY YSOs program \citep{Wojtczak2023} and on the populations of CTTSs with H$\alpha$--based $R_T$ determinations from the studies of \citet{Thanathibodee2023} and \citet{pittman2025}. 

A brief description of the GRAVITY YSO survey is provided in Sect.~\ref{sec:gravity}. The work of \citet{Thanathibodee2023} examined a sample of 20 late-K to mid-M YSOs characterized by low accretion rates. \citet{pittman2025} investigated a sample of 67 CTTSs that were targeted with the Hubble Space Telescope (HST) as part of its UV Legacy Library of Young Stars as Essential
Standards (ULLYSES) Director’s Discretionary Time program \citep{roman_duval2025}. This program was aimed at assembling a comprehensive spectral library of CTTSs to constrain their accretion physics. To address this goal, the ULLYSES targets were chosen to encompass a wide range of stellar masses and mass accretion rates. These differences in sample selection should be taken into account when comparing results of the respective $R_T$ surveys.    

The respective mass ranges of these surveys differ, so to compare their truncation radius distributions, we examined each pair of $R_T$ sets only after excluding points for which stellar mass exceeded our maximum value of 1.4~$M_\odot$, or our stellar mass was less than the minimum value of the other sample. We then performed a two-dimensional Anderson-Darling test \citep{anderson1954test} between each pair of surveys. Statistically, we find that our $R_T$ distribution is most like that of the interferometric GRAVITY sample (p-value: 0.2). In comparison to the spectroscopy-based \citet{Thanathibodee2023} and \citet{pittman2025} samples, on the other hand, our $R_T$ distribution differs at the $<1\%$ level. Nevertheless, the samples in question are small, and hence statistical robustness of these results will need to be increased with further observations.

\section{Application to the EVE mission concept} \label{sec:eve}

\subsection{Motivation} \label{sec:eve_motivation}

We still know very little about the physical conditions of protoplanetary disks' inner regions -- specifically the spatial scales of the star-disk system and the star’s rotation rate -- despite their importance for the formation and migration of Earth- and Neptune-sized planets as well as stellar spin evolution.  Indeed, the rotation rates of young stars have profound effects on their planetary systems throughout their main-sequence lifetime \citep{bastian:2020}, setting magnetic activity levels \citep{gondoin:2018}, flare energies \citep{davenport:2019}, and X-ray fluxes \citep{johnstone:2021}, which drive atmospheric erosion for planets especially on close-in orbits. 

In response to these scientific gaps, the Early eVolution Explorer \citep[EVE;][]{macgregor2025} is a NASA Astrophysics Small Explorer (SMEX) mission concept dedicated to the formation and early evolution stages of low-mass stars and their close-in planets. EVE would be NASA's first mission with simultaneous multi-band photometry in the near-UV (NUV, 200--300~nm), visible (500--900~nm), and NIR (1.1--2.0~$\mu$m).  During its 3~year primary mission, EVE would observe a minimum of 16 fields in young stellar clusters with ages 1--100~Myr for $\sim$45 days each.  EVE is ideally designed to observe these dense regions, with an observing cadence of $\sim$60~s, a large field of view $\sim$25~deg$^2$, and an angular resolution $\leq15\arcsec$.  These unique features enable EVE to address three core science objectives unachievable with current missions: (1) assess the properties of planets' primordial atmospheres, (2) establish the impact of stellar activity on planets' subsequent atmospheric evolution, and (3) determine how star-disk interaction and angular momentum regulation establish the inner limit for young migrating planets. Objectives (1) and (2) are presented in detail in \citet{zhou2026} and \citet{howard2025}, respectively. Here, we discuss how EVE's observations, coupled with the $R_T$ determination technique described in this paper, would address objective (3). 

EVE will provide new insights into what is governing the evolution of star-disk systems by determining for the first time whether winds play a key role in carrying material and angular momentum out of these systems. If this is the case, then we expect the equilibrium value of the corotation radius to move farther from the star, in relation to the truncation radius \citep[][Fig.~3]{gehrig:2025}. This in turn regulates where planet migration stops and whether stars emerge from the protoplanetary disk stage as slow or fast rotators \citep{gehrig:2023}. To address this question, EVE will determine the relative locations of $R_{CO}$ (from the NIR band) and $R_T$ (from the NUV/visible color time series) for hundreds of young accreting stars. 

Beyond testing whether or not disk winds play an important role in angular momentum transport, the measurements that EVE will make for young star-disk systems have far more extensive impacts.  The large and uniform dataset can also tackle questions such as the potential time evolution of the inner disk truncation radius \citep{gaidos:2025} and the dependence of corotation radius, and thus system architecture \citep{swain:2026}, on stellar spectral class.  The scientific community broadly agrees that the tempo and mode of disk dispersal is important to the dynamics and architecture of planetary systems \cite[e.g.,][and references therein]{alexander:2014,kimura:2025}.  The combination of EVE’s measured disk properties and the semi-major axis values of the numerous young planets that EVE will discover will shed new light on that timescale for planetary system dynamical evolution.

\subsection{Approach and simulated mission performance} \label{sec:eve_performance}

To determine the expected mission yield in terms of number of YSO detections suitable for $R_T$ and $R_{CO}$ determination, we started from a comprehensive catalog of over 1.1 million sources that are candidate members of open star clusters and associations along and around the Galactic plane, assembled from the literature as described in \citet{zhou2026}. From this initial list, we retained $\sim$35\,500 sources that are candidate disk-bearing stars based on a variety of criteria: 2MASS $J,H,K$ colors consistent with the \citet{meyer1997} PMS locus; \textit{Spitzer}/{IRAC} colors from the GLIMPSE survey \citep{churchwell2009} and/or the Spitzer Enhanced Imaging Products (SEIP; \citealp{seip}) that translate to an $\alpha_{IRAC}$ index in the range for thick to anemic disks \citep{teixeira2012}; WISE $W1,W2,W3$ photometry consistent with the Class~II locus as defined in \citet{koenig2014}; and inclusion in published YSO candidate lists based on data products from the Gaia mission, including multi-wavelength color loci using Gaia and other all-sky surveys \citep{marton2019} and variability characteristics in Gaia data \citep{marton2023}. Individual extinction ($A_V$) and distance estimates were extracted for each source of interest from Gaia DR3 \citep{gaiadr3}, when available, or extrapolated from the surrounding cluster or association members. When not available from the literature, masses ($M_\star$) were estimated from the sources' $A_V$, distance and age\footnote{The best age estimate for each source was assumed as the average age of its parent population, as extracted from the source papers of our master catalog (\citealp{zhou2026} and references therein).} estimates and TESS magnitudes, using the MIST PMS model tracks. Following \citet{gehrig:2025}, only stars with estimated masses between 0.2--1~$M_\odot$ were retained as targets of interest for EVE's objective (3).  

We anchored our yield simulations to the reference set of CTTSs that were observed as part of the ULLYSES program and its ground-based complementary campaign, PENELLOPE \citep{manara2021}. The latter provided mid- to high-resolution optical and NIR spectroscopy for accretion and stellar parameter determination, using the Very Large Telescope (VLT) instruments X-shooter, UVES and ESPRESSO. From those surveys, we retained 45 sources distributed in mass between 0.15\,$M_\odot$ and 1.25\,$M_\odot$, in \macc{} between $3\times10^{-11}$\,$M_\odot/yr$ and $4.5\times10^{-8}$\,$M_\odot/yr$, and with a 1-1.5~dex spread in $\dot{M}_{acc}$ at any given mass. For each source, we combined the HST COS+STIS and X-shooter spectra to cover the entire wavelength range from $\sim1300$\,\AA{} to 2.5\,$\mu$m. We then used these reference spectra to map  measured $u/U/B/TESS$ magnitudes to predicted fluxes in the EVE bandpasses. This step was achieved by generating, for each source in the input list, a synthetic population of 100 sources with $M_\star$, distance and $A_V$ extracted at random from a normal distribution centered around the best literature estimates for these parameters and with standard deviation equal to their respective uncertainties for distance and $A_V$, and 20\% in $M_\star$ \citep{hillenbrand2009}. For each synthetic source, we identified the reference spectrum that best matches the source's mass, scaled it to the source's distance and $A_V$\footnote{The differential reddening correction on the ULLYSES spectra was introduced by using the Python \texttt{dust\_extinction} package \citep{Gordon2024} and the Galactic interstellar dust extinction curve by \citet{gordon2023}, with a total-to-selective absorption ratio $R_V = 3.1$.}, and estimated the expected NUV+visible+NIR fluxes from the source by convolving the scaled spectrum with mock EVE passbands (error functions with a top-hat central region centered around the middle point of the wavelength ranges listed at the beginning of Sect.~\ref{sec:eve_motivation}). Synthetic magnitudes in $u/U/B/TESS$ filters were also extracted from the same scaled spectra by convolving them with the transmission curves relevant to each filter\footnote{In the $u$-band, we used both the SDSS system \citep{fukugita1996} and the SkyMapper Southern Sky Survey system \citep{onken2024}, which correspond to the two primary sources of archival photometry that we matched to our input catalog.}, as retrieved from the SVO Filter Profile Service.

For each source of interest from our input catalog, the procedure described above produced statistical distributions of synthetic fluxes in EVE passbands that correspond to a wide range of possible accretion states, as a function of synthetic $u/U/B/TESS$ magnitudes. From these distributions, we built magnitude-to-flux conversion scales by binning the synthetic magnitude ranges in steps of 0.2~mag. We matched each source's observed brightness to the closest bin in the conversion scales and assigned simulated EVE fluxes extracted at random from a triangular distribution defined by the minimum, median, and maximum of the synthetic EVE flux distribution in the selected magnitude bin. While the actual accretion state that individual EVE targets will display cannot be known beforehand, this approach ensures that a statistically robust distribution of accretion properties is assumed at the population level to evaluate target detectability and the projected mission performance. 

EVE's ability to differentiate between wind and no-wind scenarios of angular momentum regulation in YSOs rests on a homogeneous determination of $R_T$, $R_{CO}$ (hence rotation period, $P_{rot}$), and \macc{} for a sufficiently large sample of young star-disk systems ($\gtrsim$250; \citealp{gehrig:2025}). Therefore, in order to estimate the mission yield, we applied statistical cuts to the input catalog of disk-bearing sources that account for and remove the fraction of YSOs for which one or more of the parameters listed here would not be measurable:  disk-bearing stars that are not accreting, accreting stars that do not display a clear periodicity in their light curves, and periodic sources with variability dominated by transiting inner disk structures (dipper stars) rather than stellar features such as accretion spots. The corrective factor for the ratio of accreting to disk-bearing stars was calculated from the age-dependent relationships for disk and accretion fractions provided in \citet[][$\sim$0.74 at 3~Myr]{fedele2010}. Published rotation rate distributions for young star clusters monitored from space in optical light \citep{venuti2017,rebull2018,rebull2020} were used to infer the fraction of non-dipper disk-bearing stars for which a period can be measured from their light curve ($\sim$0.65), and the fraction of those for which the measured $P_{rot}$ is shorter than one third of the planned EVE stare duration (such that at least three full rotation cycles would be covered for a robust $P_{rot}$ determination; 0.99 for 45-days-long stares). 

We downsampled the input catalog by the three corrective factors discussed above, randomly selecting the sources to be removed. For the remaining sources, we calculated the projected signal-to-noise ratio (SNR) in the three EVE bandpasses. These calculations assume telescope apertures of 12.5~cm in the NUV, 20.0~cm in the visible and 18.0~cm in the NIR, and include noise contributions from sky background, readout, dark current, and electronics. We discarded as non-detections any sources with SNR that would preclude determination of the physical parameters of interest to the precisions discussed in \citet{gehrig:2025}. Finally, we investigated whether the surviving detections may be affected by blending in the EVE pixels. This assessment was anchored to the Gaia survey in the visible and the 2MASS survey in the NIR. From the catalog of either survey, we selected all point sources that fall closer than the EVE angular resolution to a disk-bearing star in EVE's input catalog. We then flagged as potential blend any EVE target star with either a brighter neighbor or a neighbor within 2~mag brightness in $G$ and/or $H$-band.  

At the end of these procedures, we estimate that EVE would provide measurements of $R_T$, $R_{CO}$ and $\dot{M}_{acc}$ for at least 375 low-mass YSOs across 16 fields, including Orion, to differentiate models of spin evolution at the 3\,$\sigma$ level. 

\subsection{Parameter determination and hypothesis testing}

For each surveyed YSO that meets the criteria in Sect.~\ref{sec:eve_performance}, the three key parameters $R_T$, $R_{CO}$ and $\dot{M}_{acc}$ would be determined from EVE data as follows.
\begin{itemize}

\item $R_T$ values would be estimated by applying the modeling approach presented in this work to EVE's NUV and visible time series.

\item $\dot{M}_{acc}$ values would be estimated by measuring the NUV flux excess due to accretion relative to the stellar photospheric emission spectrum. Non-accreting, weak-lined T Tauri stars (WTTSs) in the same young stellar populations would be used to perform a statistical calibration of the spectral emission by the stellar photosphere as a function of stellar mass or spectral type. The NIR flux measurement would serve as a reference for anchoring the photospheric flux, and the NUV--visible flux ratio relative to the expected photospheric level would provide an estimate of the accretion shock parameters. Bolometric corrections to convert EVE's NUV flux excess into total accretion luminosities $L_{acc}$ would be derived by using the \citet{robinson2019} accretion shock models, scaled to fit the NUV+visible excess emission of EVE targets. In turn, $L_{acc}$, $\dot{M}_{acc}$ and $R_T$ are connected by the equation $L_{acc} = \frac{G M_\star \dot{M}_{acc}}{R_\star}\left(1-\frac{R_\star}{R_T} \right)$ \citep{gullbring1998}, where G is the gravitational constant.

\item $R_{CO}$ values would be calculated from the measured stellar $P_{rot}$ by assuming inner disk Keplerian rotation, as $R_{CO} = \sqrt[3]{G M_\star \left(\frac{P_{rot}}{2 \pi}\right)^2}$. $P_{rot}$ values would be determined from the rotational modulation effect induced by starspots on EVE's NIR light curves.

\end{itemize}

While rotation rates for young stars are routinely measured from optical broadband light curves, the accretion shock can make a contribution comparable to the stellar photosphere to the total flux measured at visible wavelengths \citep[e.g.,][]{pittman2022}. Because accretion shocks can display significant evolution on timescales different from the stellar rotation cycle \citep[e.g.,][]{espaillat2021}, this may lead to both a lower fraction of recoverable rotation rates and more uncertain or erroneous $P_{rot}$ estimates in YSOs accreting at typical rates $\dot{M}_{acc} \sim 1-5\times 10^{-8}\ M_\odot/yr$.

To quantitatively assess the advantage of NIR over optical curves to rotation period determination, we compared published $P_{rot}$ distributions for disk-bearing stars in the Orion Nebula Cluster that were extracted from optical light curves \citep[$I$-band;][]{herbst2002,rodriguez_ledesma2009} to results obtained in the NIR \citep[$J/H/K$-bands;][]{rice2015}. Based on a sample of over 300 disk-bearing stars in common between the optical and NIR studies, we found that the fraction of stars that appear periodic in the NIR ($f_{NIR}$) is 15\% larger than in the optical. Disk-bearing YSOs are known to transition between periodic and stochastic variability behaviors over timescales of years or less \citep{sousa2016}, and therefore part of the discrepancy between $f_{NIR}$ and $f_{optical}$ may be ascribed to the non-simultaneity of the relevant surveys. However, the fact that both optical surveys independently reported fewer periodic behaviors than the NIR survey for YSOs in common suggests an underlying difference in the mix of phenomena that lead to the observed variability behaviors in the two wavelength regimes (see also discussion by \citealp{rice2015}). 

We further assessed the difference in the optical-derived vs. NIR-derived $P_{rot}$ values for sources that were reported as periodic by either \citet{herbst2002} or \citet{rodriguez_ledesma2009} and \citet{rice2015}. Stochastic fluctuations in the measured periods due to photometric uncertainties would lead to a normal distribution of $P_{rot}$(optical)--$P_{rot}$(NIR) residues centered on 0. However, in both samples we found a surplus of sources with inferred $P_{rot}$(optical) shorter than the corresponding $P_{rot}$(NIR). This result is at least qualitatively consistent with a scenario where rapidly-evolving accretion shocks, driven by unstable accretion streams on very short inner disk timescales \citep{romanova2009},  contribute a significant fraction of the measured flux at optical wavelengths, thus lowering the apparent $P_{rot}$. We estimate that this bias in the $P_{rot}$ determinations may affect 25\%--30\% of disk-bearing YSOs. EVE's NIR time series data, with a bandpass designed to cover the 2MASS $J+H$ filters (where the accretion shock emission is minimal and the emission from dust in the inner disk is negligible), would enable circumventing these issues and provide $P_{rot}$ measurements that are independent from, and simultaneous to, the NUV--visible accretion tracers that would yield $R_T$ and $\dot{M}_{acc}$.

The spin population model by \citet{gehrig:2025} would be used to generate synthetic populations of CTTSs that reproduce the ($M_\star$, $\dot{M}_{acc}$, $P_{rot}$) composition of the observed sample and are assumed to be in rotational equilibrium with or without the assistance of winds (see their Fig.~5). The observed distributions of $R_T$, $R_{CO}$ and $\dot{M}_{acc}$ would then be tested against the synthetic populations, using both non-parametric k-sample statistical tests and a KDE approach, to determine the most plausible physical scenario behind spin regulation in YSOs.

\section{Conclusions} \label{sec:conclusions}

Measurements of the inner disk truncation radius around young stars are key to revealing the physical conditions of star-disk mass and angular momentum transfer and spin equilibrium. Homogeneous, statistical surveys of $R_T$, coupled with $R_{CO}$ determinations for entire populations of young stars, can help distinguish among models of angular momentum regulation in YSOs, and they can shed new light on scenarios of planet formation and dynamical orbit evolution in the inner disk. Several methods have been previously proposed to estimate $R_T$ in YSOs, from spatially resolving the inner disk magnetospheric accretion region via NIR interferometry, to modeling line emission by the accelerated material in the accretion columns. However, the collections of $R_T$ measurements that are currently available based on those approaches are limited in number, and their inhomogeneous nature prevents a reliable assessment of underlying physical trends that may be masked by model-dependent effects.

In this work, we have proposed a new method for determining $R_T$ in young stars that has the potential to provide an order of magnitude increase in the number of stars for which homogeneous measurements of $R_T$ can be achieved. Our method leverages the color dependence of the accretion shock emission to infer the distance traveled by the accretion stream from its launching point at the inner disk edge to the star. The color and luminosity of the flux excess produced by accretion at NUV and optical wavelengths depend on the kinetic energy flux of the accretion flow, which in turn depends on its density and free-fall velocity, largely determined by $R_T$. We used the \citet{robinson2017,robinson2021} models to simulate the distributions of multi-color time-series photometry in color-magnitude space as a function of $M_\star$ and $R_T$, modulated by stellar inclination. We developed a KDE-based approach to fit these theoretical predictions to multi-band time-domain observations of accreting stars and identify the $R_T$ value of the model that best reproduces the observed color distribution. We tested our method on representative YSOs with magnetospheric size estimates from GRAVITY interferometry and contemporaneous LCOGT $u,g,r,i$ monitoring suitable for application of our approach. We showed that results from our model fits for these test cases agree well with interferometric proxies for the inner gas disk size.

We used our method to estimate the $R_T$ of 26 disk-bearing YSOs in the Taurus and Upper Scorpius star-forming regions. We derived an $R_T$ measurement for 17 of our targets; most of the sources for which an $R_T$ solution could not be extracted are more evolved YSOs with weaker accretion signatures. Our results show a predominance of smaller $R_T$ values than typically assumed ($<4\ R_\star$), particularly at the lowest stellar masses ($M_\star \leq 0.3\ M_\odot$).  Our distribution of values is statistically consistent with that of the interferometric studies in the literature. However, we note that our sample of stars is skewed towards somewhat lower stellar masses. Thus, our photometric approach can complement other existing techniques and help expand $R_T$ surveys down to regions of the stellar parameter space that may not be accessible with other diagnostics.

Finally, we discussed how our technique can be applied to the high-cadence NUV/visible/NIR light curves that would be generated by the proposed NASA SMEX mission EVE to infer homogeneous $R_T$ measurements for several hundred young star-disk systems. When combined with the simultaneous measurements of accretion rates $\dot{M}_{acc}$ and corotation radii $R_{CO}$ that the mission would also provide for the same sources, this unprecedented survey of inner disk properties would enable the first stringent test of angular momentum evolution theories in YSOs. Together with EVE's other core science objectives of determining the energy environments and primordial atmospheric composition of small, close-in planets, these measurements will provide the most detailed view yet of how the inner disk conditions shape the early architectures of planetary systems and influence their subsequent evolution.   

\begin{acknowledgments}
This research benefited from LV, CER, AMC, and EG’s participation in the 2026 workshop “Exploring Planetary Systems in the Era of Time-domain Astronomy”, hosted by the Institute for Astronomy at the University of Hawai’i and supported by award 2106927 from the National Science Foundation’s
Astronomy \& Astrophysics Research Grants program. This work was partly supported by the National Aeronautics and Space Administration (NASA) under grant No. 80NSSC25K7884 issued through the NNH24ZDA001N Astrophysics Data Analysis Program (ADAP). EG also acknowledges support from the NASA ADAP award 80NSSC19K0587. This work was carried out in part at the Jet Propulsion Laboratory, California Institute of Technology, under contract 80NM00018D0004 with NASA. This research makes use of Lightkurve, a Python package for Kepler and TESS data analysis \citep{2018ascl.soft12013L}. This work also makes use of observations from the Las Cumbres Observatory global telescope network. This paper includes data collected by the TESS mission. Funding for the TESS mission is provided by the NASA's Science Mission Directorate. This work presents results from the European Space Agency (ESA) space mission Gaia. Gaia data are being processed by the Gaia Data Processing and Analysis Consortium (DPAC). Funding for the DPAC is provided by national institutions, in particular the institutions participating in the Gaia MultiLateral Agreement (MLA). This publication makes use of data products from the Wide-field Infrared Survey Explorer, which is a joint project of the University of California, Los Angeles, and the Jet Propulsion Laboratory/California Institute of Technology, funded by NASA. 
\end{acknowledgments}

\vspace{5mm}
\facilities{TESS, LCOGT}

\software{Matplotlib \citep{matplotlib}, NumPy \citep{numpy}, SciPy \citep{scipy}, TOPCAT \citep{topcat}, R \citep{R}}

\bibliography{references}{}
\bibliographystyle{aasjournal}

\end{document}